\documentclass[structabstract]{aa} 
\usepackage{graphicx}
\usepackage{txfonts}
\usepackage[]{natbib}
\usepackage{array}

\begin{document}

\title{Galaxy clusters in presence of dark energy: a kinetic approach}

\author{M. Merafina$^1$, G.S. Bisnovatyi-Kogan$^{2,3}$, M. Donnari$^{1,4}$}
\institute{$^1$ Department of Physics, University of Rome ``La Sapienza", Piazzale Aldo Moro 2, I-00185 Rome, Italy\\
$^2$ Space Research Institute, Russian Academy of Sciences, Moscow, Russia\\
$^3$ National Research Nuclear University MEPhI, Moscow, Russia\\
$^4$ Department of Physics, University of Rome ``Tor Vergata", Via della Ricerca Scientifica 1, I-00133 Rome, Italy\\
\\
\email{marco.merafina@roma1.infn.it; gkogan@iki.rssi.ru; martina.donnari@roma1.infn.it}}

\titlerunning{Galaxy clusters in presence of dark energy: a kinetic approach}
\authorrunning{Merafina et al.}

\abstract{The external regions of galaxy clusters may be under strong influence of the dark energy, discovered by observations of the SN Ia at redshift $z<1$.}{The presence of the dark energy in the gravitational equilibrium equation, with the Einstein $\Lambda$ term, contrasts the gravity, making the equilibrium configuration more extended in radius.}{In this paper we derive and solve the kinetic equation for an equilibrium configuration in presence of the dark energy, by considering the Newtonian regime, being the observed velocities of the galaxies inside a cluster largely smaller than the light velocity.}{The presence of the dark energy in the gravitational equilibrium equation leads to wide regions in the $W_0$-  $\rho_\Lambda$ diagram where the equilibrium solutions are not permitted, due to the prevalence of the effects of the dark energy on the gravity.}{}


\maketitle

\section{Introduction}

It was shown by \cite{2001Chernin,2008Chernin} that outer parts of galaxy clusters (GC) may be under strong influence of the dark energy (DE), discovered by observations of SN Ia at redshift $z\le 1$ \citep{1998Riess,1999Perlmutter}, and in the spectrum of fluctuations of CMB radiation \citep[see e.g.][]{2003Spergel,2004Tegmark}. Equilibrium solutions for polytropic configurations in presence of DE have been obtained in papers of \cite{2006Balaguera,2007Balaguera} and \cite{2012Merafina}. Here we derive a Boltzmann-Vlasov kinetic equation in presence of DE, in Newtonian gravity, and obtain its solutions. These solutions generalize the ones, obtained by \cite{1993Bisnovatyi,1998Bisnovatyi} for the kinetic equation without DE. Here we consider the problem in Newtonian approximation, because the observed chaotic velocities of galaxies inside a cluster are much less than the light velocity $v_{gc}\ll c$.

  The general relativistic solution in presence of DE could be applied for equilibrium configurations of point masses from some exotic particles, interacting only gravitationally. On early stages of the universe expansion, before and during the inflation stage, these particles may form gravitationally bound configurations which collapse during the inflation, when antigravity decreases. As a result of this collapse such hypothetical objects may be transformed into primordial black holes appearing after the end of inflation. The relativistic kinetic equation, and its solutions in presence of DE will be considered elsewhere.

\section{Newtonian approximation in description of DE}

The substance, which is called now as DE was first introduced by Einstein (1918) for a stationary universe, in the form of cosmological constant $\Lambda$, during his unsuccessful attempts to construct a solution for a stationary universe. Soon after de Sitter (1917) had shown, that in presence of $\Lambda$ the solution for an empty space describes an exponential expansion. \cite{1922Friedmann,1924Friedmann} was the first, who had obtained exact solutions for the expanding universe containing matter, in presence of the cosmological constant $\Lambda$. Another exact solution for the metric in presence of $\Lambda$, around the gravitating point mass,  was obtained by \cite{Carter}.
This solution is a direct generalization of the Schwarzschild solution for a black hole (BH) in vacuum with a metric
\begin{equation}
ds^2=g_{00}c^2 dt^2-g_{11}dr^2-r^2(d\theta^2+\sin^2\theta d\varphi^2)\ ,
 \label{eq1}
\end{equation}
has a form
\begin{equation}
 \label{eq2}
 g_{00}=\frac{1}{g_{11}}=1-\frac{2GM}{c^2 r}-\frac{\Lambda r^2}{3}
  =1-\frac{2GM}{c^2 r}-\frac{8\pi G \rho_\Lambda r^2}{3c^2}\ ,
\end{equation}
where the density of DE $\rho_\Lambda$ is connected with $\Lambda$ as
\begin{equation}
\rho_\Lambda=\frac{\Lambda c^2}{8\pi G}\ .
 \label{eq3}
\end{equation}
A transition to the Newtonian limit, where DE is described by the antigravity force in vacuum was done by \cite{2008Chernin}. In the limit of a weak gravity ($v^2 \ll c^2$, $GM/r \ll c^2$) the metric coefficients are connected with a gravitational potential $\Phi_g$ as \citep{Landau}
\begin{equation}
 g_{00}^{1/2}=1+\frac{\Phi_g}{c^2}\ .
 \label{eq4}
\end{equation}
Then, using the Eq.(\ref{eq4}) and the Eq.(\ref{eq2}) at $\Lambda=0$, we obtain the expression for the Newtonian potential $\Phi_g$ and the Newtonian gravity force acting on the unit mass $F_g$
\begin{equation}
 \Phi_g=-\frac{GM}{r}, \quad F_g=-\frac{d\Phi_g}{dr}=-\frac{GM}{r^2}\ .
 \label{eq5}
\end{equation}
For the Schwarzschild-de Sitter metric (\ref{eq2}) we have in the Newtonian limit
\begin{equation}
\Phi=-\frac{GM}{r}-\frac{4\pi G\rho_\Lambda r^2}{3}\ ,\; F = F_{g}+F_{\Lambda}=-\frac{GM}{r^2} +\frac{8\pi G\rho_\Lambda r}{3}.
 \label{eq6}
\end{equation}
So, the cosmological constant create in a vacuum a repulsive (antigravity) force between a BH and a test particle, which increases linearly with a distance between them. The normalization of the potential here is chosen so that $\Phi_g=0$ at $r=\infty$, and $\Phi_\Lambda=0$ at $r=0$.

Let us consider now the equilibrium of a self-gravitating object in presence of DE. In general relativity the equations describing the equilibrium in a spherically symmetric configuration in vacuum (without DE) had been derived by \cite{1939Oppenheimer}
\begin{equation}
\frac{dP}{dr}=-\frac{G(\rho c^2+P)(M_r c^2 +4\pi P r^3)}{r^2c^4 -2GM_r rc^2}
 \label{eq7}
\end{equation}
\begin{displaymath}
\frac{dM_r}{dr}=4\pi \rho r^2\ .
\end{displaymath}
Here $\rho$, $P$ are the total density and total pressure of the matter, and $M_r$ is the total (gravitating) mass, including a gravitational binding energy, inside a radius $r$ in the Schwarzschild-like metric
\begin{equation}
ds^2=e^\nu c^2 dt^2-e^\lambda dr^2-r^2(d\theta^2+\sin^2\theta d\varphi^2)\ ,
 \label{eq8}
\end{equation}
\begin{equation}
e^\lambda=\left(1-\frac{2GM_r}{r c^2}\right)^{-1}\ ,
 \label{eq9}
\end{equation}
\begin{equation}
e^\nu=\exp\left(2\int_r^\infty \frac{dP/dr}{P+\rho c^2}dr\right)\ .
 \label{eq10}
\end{equation}
In presence of DE, $\rho$ and $P$ are represented as
\begin{equation}
\rho=\rho_m +\rho_\Lambda=\rho_m+\frac{\Lambda c^2}{8\pi G}\ ,\quad P=P_m+P_\Lambda=P_m-\frac{\Lambda c^4}{8\pi G}\ .
 \label{eq11}
\end{equation}
Let us consider a Newtonian limit when
\begin{equation}
P_m \ll \rho_m\, c^2, \quad r \gg \frac{2GM_r^{(m)}}{c^2}\ .
 \label{eq12}
\end{equation}
Here we used a definition $M_r^{(m)}=4\pi\int_0^r \rho_m r^2 dr$. In the Newtonian limit we have from Eq.(\ref{eq7})
\begin{equation}
\frac{dP}{dr}=-\frac{\rho_m \left(3GM_r^{(m)}-\Lambda c^2 r^3\right)}{r^2 \left(3 - \Lambda r^2\right)}\ .
 \label{eq13}
\end{equation}
Let us estimate the last term in the denominator. For existence of an equilibrium configuration with a finite radius we need a positive sign of the numerator, so using the conditions (\ref{eq12}) we have
\begin{displaymath}
\Lambda r^2 < \frac{3GM_r^{(m)}}{rc^2} \ll 1\ .
\end{displaymath}
Therefore, in the denominator we have $\Lambda r^2 \ll 3$ and we can neglet the term with $\Lambda$. So in the Newtonian approximation, in presence of DE, we obtain the following equilibrium equation
\begin{equation}
 \frac{dP}{dr}=-\rho_m\left(\frac{GM_r^{(m)}}{r^2}-\frac{\Lambda c^2 r}{3}\right)=-\rho_m\left(\frac{GM_r^{(m)}}{r^2}-\frac{8\pi G \rho_\Lambda r}{3}\right)\ ,
 \label{eq14}
\end{equation}
with $\rho_\Lambda$ given by the definition (\ref{eq3}), which was used without derivation by \cite{2012Merafina}. 
On the other hand, we can write the Poisson equation for the gravity of the matter together with the hydrostatic equilibrium equation
\begin{equation}
\nabla^2\Phi_g=4\pi G\rho_m \ , \quad \frac{{\bf{\nabla}} P}{\rho_m}=- {\bf{\nabla}} \Phi_g - {\bf{\nabla}} \Phi_\Lambda \ , 
\label{eq15}
\end{equation}
and then, the potential created by DE in the vacuum, taking into account that $P=P_m+P_{\Lambda}$, satisfies the Poisson equation
\begin{equation}
\nabla^2\Phi_\Lambda =-8\pi G \rho_\Lambda, \quad \rho_\Lambda=\frac{\Lambda c^2}{8\pi G}\ .
 \label{eq16}
\end{equation}
This equation, together with the Poisson equation for the gravity of the matter gives a full description of a static gaseous equilibrium configuration in presence of DE. Similarly we can write the hydrodynamic Euler equation in presence of DE as
\begin{equation}
 \frac{\partial{\bf {v}}}{\partial t}+ ({\bf {v}} \cdot {\bf{\nabla}}){\bf{v}} + \frac{{\bf{\nabla}} P}{\rho_m}= -{\bf {\nabla}} \Phi_g-{\bf{\nabla}} \Phi_\Lambda\ .
 \label{eq17}
\end{equation}

\section{Kinetic equation for self-gravitating cluster in presence of DE}

The kinetic Boltzmann-Vlasov equation for a distribution function $f$ of non-collisional gravitating points of equal mass $m$, in spherical coordinates $(r,\theta,\varphi)$ is written as
\begin{displaymath}
\frac{\partial f}{\partial t}+v_r \frac{\partial f}{\partial r}+\frac{v_\theta}{r} \frac{\partial f}{\partial \theta}+ \frac{v_\varphi}{r \sin\theta} \frac{\partial f}{\partial \varphi}\,+
\end{displaymath}
\begin{displaymath}
+\left(\frac{v_\theta^2+v_\varphi^2}{r}-\frac{\partial \Phi}{\partial r}\right) \frac{\partial f}{\partial v_r}\,+ \left(-\frac{v_r v_\theta}{r}+\frac{\cot\theta\, v_\varphi^2}{r}-\frac{1}{r}\frac{\partial \Phi}{\partial \theta}\right)\frac{\partial f}{\partial v_\theta}\,+ 
\end{displaymath}
\begin{equation}
+\left(-\frac{v_r v_\varphi}{r}-\frac{\cot\theta\, v_\varphi v_\theta}{r}-\frac{1}{r\sin\theta}\frac{\partial \Phi}{\partial \varphi}\right)\frac{\partial f}{\partial v_\varphi}=0 \ ,
 \label{eq18}
\end{equation}
where, in presence of DE, we have $\Phi=\Phi_g\,+\,\Phi_\Lambda$.  In a spherically symmetric stationary cluster, we have $\partial\Phi / \partial t=0$ and $\Phi=\Phi (r)$. Moreover, the kinetic equation (\ref{eq18}) has four first integrals, written in Cartesian coordinates ($x,y,z$) as
\begin{displaymath}
\frac{E}{m}=\frac{1}{2}(v_x^2+v_y^2+v_z^2)+\Phi\ ,
\end{displaymath}
\begin{equation}
\frac{L_x}{m}=y\,v_z-z\,v_y\ ,\quad
\frac{L_y}{m}=z\,v_x-x\,v_z\ ,\quad
\frac{L_z}{m}=x\,v_y-y\,v_x\ .
 \label{eq19}
\end{equation}
In spherical coordinates, these integrals can be expressed by
\begin{displaymath}
\frac{E}{m}=\frac{1}{2}(v_r^2+v_\theta^2+v_\varphi^2)+\Phi\ ,\quad
\frac{L_x}{m}=-r\,v_\theta\sin\varphi-r\,v_\varphi\cos\theta\cos\varphi\ ,
\end{displaymath}
\begin{equation}
\frac{L_y}{m}=r\,v_\theta\cos\varphi-r\,v_\varphi\sin\varphi\cos\theta\ ,\quad \frac{L_z}{m}=r\,v_\varphi\sin\theta\ ,
 \label{eq19a}
\end{equation}
where $E$ and $L_i$ $(i=x,y,z)$ are the energy and the projection of the angular momentum on the corresponding axis, respectively. From the last three integrals follows the conservation of the absolute value of the angular momentum $L$, written in the form
\begin{equation}
\frac{L^2}{m^2}=r^2(v_\theta^2+v_\varphi^2)\ .
\end{equation}
Then, the solution of the kinetic equation (\ref{eq18}) is an arbitrary function of the first integrals (\ref{eq19a}). We restrict ourselves to an isotropic distribution function $f(E)$. For a uniform DE, a normalization of its energy at $r=\infty$ is not possible, therefore we choose $\Phi_\Lambda=0$ at $r=0$ as the most convenient one \citep{2012Merafina}. Thus, from Eqs.(\ref{eq15}), we have
\begin{equation}
\Phi_\Lambda=-\frac{4\pi G}{3}\rho_\Lambda r^2=-\frac{\Lambda c^2}{6}r^2
\ .
 \label{eq20}
\end{equation}
Following \cite{1965Zeldovich}, \cite{1993Bisnovatyi,1998Bisnovatyi}, we consider a Maxwell-Boltzmann distribution function with a cut-off

\begin{equation}
\left\{\begin{array}{ll}
f=B e^{-E/T} & \mbox{\qquad for\quad} E \leq E_{cut}\\
& \\
f=0 & \mbox{\qquad for\quad} E > E_{cut}\ ,
\end{array}
\right.
 \label{eq21}
 \end{equation}
where the cutoff energy $E_{cut}$ is given by
\begin{equation}
E_{cut}=-\frac{\alpha T}{2} 
 \label{eq21a}
\end{equation}
and $\alpha$ is the so called cutoff parameter, while $T$ is the temperature in energy units. The total energy is 
\begin{equation}
E\,=\,\frac{mv^2}{2}+m\Phi=\frac{mv^2}{2}+m\Phi_g-\frac{m\Lambda c^2 r^2}{6}\ ,
\end{equation}
where the total potential $\Phi$ and the velocity $v$ are given by
\begin{equation}
\Phi = \Phi_g +\Phi_{\Lambda}\quad\quad {\rm and}\quad\quad v=(v_r^2+v_\theta^2+v_\varphi^2)^{1/2}\ .
\end{equation}

The constant $B$ in the first of the Eqs.(\ref{eq21}) depends on the total potential $\Phi$ and therefore it is not the same for each model. In order to consider a unique distribution function for all the equilibrium 
configurations, following \cite{1989Merafina}, we must choose a different
normalization by introducing a new constant $A$ connected with $B$ through the following relation\footnote{\,Eq.(\ref{eq21z}) is not arbitrary but justified by considerations of statistical mechanics, being 
$E=\rm{constant}$ along the motion of each single component of mass $m$ and taking into account the presence of the chemical potential $\mu$ in the constant $B$, where $\mu+m\Phi=\rm{constant}$ along the radial coordinate 
$r$.}
\begin{equation}
B=Ae^{m\Phi_R/T} \ ,
 \label{eq21z}
\end{equation}
with $\Phi_R$ the value of the total potential $\Phi$ at $r=R$. In this way, the expression of the distribution function (\ref{eq21}) for $E\leq E_{cut}$ becomes
\begin{equation}
f=A\,{\rm{exp}} \left[\frac{m\Phi_R}{T}-\frac{mv^2}{2T}-\frac{m}{T}\left(\Phi_g- \frac{\Lambda c^2 r^2}{6}\right)\right] \ .
 \label{eq21b}
\end{equation}
The maximum kinetic energy $\epsilon_c$ is connected with the potential $\Phi$ by the relation
\begin{equation}
\epsilon_c=m(\Phi_R-\Phi) \ .
 \label{eq21c}
 \end{equation}
Then the distribution function can be rewritten as
\begin{equation}
f=A\,e^{-(\epsilon-\epsilon_c)/T} \ ,
 \label{eq22}
\end{equation}
where $\epsilon=mv^2/2$ is the kinetic energy of the single point mass.

The Poisson equation (\ref{eq15}) in a spherical symmetry referred to gravitational field is given by
\begin{equation}
\frac{1}{r^2}\frac{d}{dr}\left(r^2\frac{d\Phi_g}{dr}\right)=4\pi  G\rho_m
\ ,
 \label{eq23}
\end{equation}
with the boundary conditions $\Phi_g (0)=\Phi_{g0}$ and $\Phi_g' (0)=0$. The matter density can be expressed as
\begin{equation}
 \rho_m=4\pi m\int_0^{p_{max}} f p^2 dp\ ,\quad\quad {\rm with}\quad\quad p=mv\ ,
 \label{eq24}
\end{equation}
where the expression of the maximum momentum $p_{max}$ is given by
\begin{displaymath}
p_{max}=\sqrt{2m\left(-m\Phi-\frac{\alpha T}{2}\right)} =\sqrt{2m\left(-m\Phi_g+\frac{m\Lambda c^2 r^2}{6}-\frac{\alpha T}{2}\right)}\ .
\end{displaymath}
The cluster with a finite radius is possible only when the following condition is satisfied 
\begin{displaymath}
\frac{\alpha T}{2m}<-\Phi_{max}\ , \quad\quad {\rm with}\quad\quad \Phi_{max}<0\ .
\end{displaymath}
Then, by using the form of the distribution given in Eq.(\ref{eq22}), we can finally rewrite the matter density as
\begin{equation}
 \rho_m=4\sqrt{2}\pi Am^{5/2}\int_0^{\epsilon_c} e^{-(\epsilon -\epsilon_c) /T} \sqrt{\epsilon}\, d\epsilon\ ,\quad {\rm where} \quad \epsilon_c=\frac{p^{2}_{max}}{2m}\ .
  \label{eq24a}
\end{equation}
Introducing dimensionless variables
\begin{equation}
W=\frac{p^2_{max}}{2mT}=\frac{\epsilon_c}{T}\quad\quad {\rm and}\quad\quad x=\frac{p^2}{2mT}=\frac{\epsilon}{T}\ ,
 \label{eq25}
\end{equation}
we obtain $W=m(\Phi_R -\Phi)/T$ and the expression of matter density $\rho_m$ becomes
\begin{equation}
 \rho_m=4\sqrt{2}\pi Am^{5/2}T^{3/2}\int_0^W e^{W-x} \sqrt{x} dx\ ,
  \label{eq25a}
\end{equation}
where, as usual, at $W=0$ we have $\rho_m=0$, being $\Phi =\Phi_R$\,.

From the Poisson equation (\ref{eq23}) we can deduce the equation describing the structure of the Newtonian configurations in presence of DE by considering also the potential $\Phi_{\Lambda}$. In fact, inserting the expression of the gravitational potential $\Phi_g = \Phi - \Phi_{\Lambda}$ in Eq.(\ref{eq23}) and using the Eq.(\ref{eq20}), we obtain 
\begin{equation}
\frac{1}{r^2}\frac{d}{dr}\left(r^2\frac{d\Phi}{dr}\right)=4\pi G\rho_m - \Lambda c^2 \ ,
 \label{eq26}
\end{equation}
where the potential $\Phi$ now includes all the contributions. Then, by considering the first relation in Eq.(\ref{eq11}), the equilibrium equation becomes
\begin{equation}
\frac{1}{r^2}\frac{d}{dr}\left(r^2\frac{d\Phi}{dr}\right)=4\pi G (\rho_m - 2\rho_{\Lambda})\ .
 \label{eq26a}
\end{equation}
Now, we have to consider the boundary conditions for the potential $\Phi$ with respect to ones given for the potential $\Phi_g$ in the Eq.(\ref{eq23}). Starting from Eq.(\ref{eq20}), we can write
\begin{equation}
\Phi =\Phi_g -\frac{\Lambda c^2}{6}r^2\quad\quad {\rm and}\quad\quad
\Phi' =\Phi_g' -\frac{\Lambda c^2}{3}r 
 \label{eq27}
\end{equation}
and therefore, for $r=0$, we have $\Phi(0)=\Phi_{g0}$ and $\Phi'(0)=\Phi_g' (0)=0$.

In order to write the dimensionless form of the equilibrium equation, we can express the radial coordinate as $r=\eta\hat r$ and, using the definition $W=m(\Phi_R -\Phi)/T$, the equilibrium equation can be rewritten as
\begin{equation}
\frac{1}{\hat r^2}\frac{d}{d\hat r}\left(\hat r^2\frac{dW}{d\hat r}\right)=-\frac{4\pi Gm\eta^2}{T}(\rho_m - 2\rho_{\Lambda})\ .
 \label{eq27a}
\end{equation} 
In the same way, following \cite{1989Merafina}, we can introduce the expression of dimensionless densities by defining the following quantities
\begin{equation}
\rho_m=\frac{\sigma^2}{G\eta^2}\hat\rho_m\quad\quad {\rm and}\quad\quad
\rho_{\Lambda}=\frac{\sigma^2}{G\eta^2}\hat\rho_{\Lambda}\ ,
 \label{eq28}
\end{equation}
where $\sigma^2=2T/m$. Thus, the dimensionless form of the equilibrium equation will be given by
\begin{equation}
\frac{1}{\hat r^2}\frac{d}{d\hat r}\left(\hat r^2\frac{dW}{d\hat r}\right)=-8\pi(\hat\rho_m - 2\hat\rho_{\Lambda})\ ,
 \label{eq29}
\end{equation}
with the boundary conditions $W(0)=W_0$ and $W'(0)=0$. Moreover, it is important to note that the relation $\hat\rho_{m0} > 2\hat\rho_{\Lambda}$ must be satisfied at the center of the equilibrium configuration in order to obtain the condition of initial decreasing density $W''(0)<0$. However, this is a necessary but not sufficient condition for the existence of the equilibrium solution, because the presence of the DE can make possible to reach conditions of increasing density ($W'>0$) at other values of the radial coordinate.
 
It remains to define the expression of the dimensional quantity $\eta$. For  getting the result, we can use the relations (\ref{eq25a}) and (\ref{eq11}) for the densities $\rho_m$ and $\rho_{\Lambda}$, respectively, and compare them with the definitions (\ref{eq28}). We obtain
\begin{equation}
\eta=(Am^4G\sigma)^{-1/2}\ ,
 \label{eq30}
\end{equation}
with
\begin{equation}
\hat\rho_m= 2\pi\int_0^W e^{W-x}\sqrt{x}dx \quad\quad {\rm and}\quad\quad \hat\rho_{\Lambda}=\frac{\Lambda\eta^2 c^2}{8\pi\sigma^2}\ ,
 \label{eq30a}
\end{equation}
where $\hat\rho_{\Lambda}$ is given by the value of $\Lambda$. The total mass $M^{(m)}$ at radius $R$ is given by
\begin{equation}
M^{(m)}=4\pi\int_0^R \rho_m r^2 dr=\frac{\sigma^2 \eta}{G}\int_0^{\hat R}
4\pi\hat\rho_m \hat r^2 d\hat r\ ,
 \label{eq30b}
\end{equation}
where
\begin{equation}
\hat M^{(m)}=\int_0^{\hat R} 4\pi\hat\rho_m \hat r^2 d\hat r \quad\quad {\rm and}\quad\quad M^{(m)}=\frac{\sigma^2 \eta}{G}\hat M^{(m)} \ .
 \label{eq30c}
\end{equation}

Finally, in order to make explicit the dependence of the dimensional quantities on the velocity $\sigma$, we can introduce the quantity
\begin{equation}
\zeta = \eta\,\sigma^{1/2}=(Am^4 G)^{-1/2} 
 \label{eq30d}
\end{equation}
and the dimensional quantities can be rewritten as
\begin{equation}
\rho_m =\frac{\sigma^3}{G\zeta^2}\hat\rho_m \quad\quad {\rm and}\quad\quad \rho_{\Lambda} =\frac{\sigma^3}{G\zeta^2}\hat\rho_{\Lambda}
 \label{eq30e}
\end{equation}
and
\begin{equation}
M^{(m)}=\frac{\sigma^{3/2}\zeta}{G}\hat M^{(m)} \quad\quad {\rm and}\quad\quad R=\frac{\zeta}{\sigma^{1/2}}\hat R \ .
 \label{eq30f}
\end{equation}

Turning to the condition (\ref{eq21a}) on the energy $E_{cut}\,$, we can express the cutoff parameter $\alpha$ by using the condition at the edge of the configuration
\begin{equation}
 \frac{\alpha}{2}=-\frac{m\Phi_R}{T}=-\frac{m}{T}(\Phi_g +\Phi_{\Lambda})_{r=R}\ .
 \label{eq30g}
\end{equation}
Thus, being $\Phi_g (R)=-GM^{(m)}/R$ and $\Phi_{\Lambda} (R)=-\Lambda c^2 R^2 /6$, we obtain
\begin{equation}
 \alpha=\frac{2GmM^{(m)}}{RT}+\frac{m\Lambda c^2 R^2}{3T}
 \label{eq30h}
\end{equation}
and, finally, by using dimensionless quantities (\ref{eq30e}), (\ref{eq30f}) and the relation (\ref{eq3}), we have
\begin{equation}
 \alpha=\frac{4\hat M^{(m)}}{\hat R}\left({1+\frac{4\pi\hat \rho_{\Lambda}\hat R^3}{3\hat M^{(m)}}}\right)\ .
 \label{eq30i}
\end{equation}

For small values of the cutoff parameter $\alpha$, maintaining a finite value of $\alpha\,T$ which corresponds to high values of the temperature 
$T$, the distribution function (\ref{eq21}) may be taken as a constant \citep{1998Bisnovatyi}. Then, the solutions exist only for small values of $W_0$ and $\hat\rho_{\Lambda}$, assuming a more simplified form which converges to a limiting sequence. In the limit of $W\to 0$, the dimensionless density $\rho_m$ can be expressed as
\begin{equation}
\hat\rho_m= 2\pi\int_0^W \sqrt{x}dx = \frac{4\pi}{3} W^{3/2}\ ,
 \label{eq30j}
\end{equation}
whereas the equilibrium equation (\ref{eq29}) becomes
\begin{equation}
\frac{1}{\hat r^2}\frac{d}{d\hat r}\left(\hat r^2\frac{dW}{d\hat r}\right)=-\frac{32\pi^2}{3}W^{3/2}+16\pi\hat\rho_{\Lambda}\ .
 \label{eq31}
\end{equation}
Expressing in dimensional terms, the density can be written as
\begin{equation}
\rho_m =\frac{4\pi}{3}\frac{\sigma^3}{G\zeta^2}W^{3/2}=\rho_p W^{3/2}\ ,  
 \label{eq31a}
\end{equation}
where
\begin{equation}
\rho_p =\frac{4\pi}{3}\frac{\sigma^3}{G\zeta^2}\ .   
 \label{eq31b}
\end{equation}
Moreover, by imposing a change of radial coordinate from $\hat r$ to $y$ for which
\begin{equation}
\hat r =y\,\left({\frac{3}{32\pi^2}}\right)^{1/2}\ ,  
 \label{eq31c}
\end{equation}
the dimensionless equilibrium equation can be rewritten as
\begin{equation}
\frac{1}{y^2}\frac{d}{dy}\left(y^2\frac{dW}{dy}\right)=-W^{3/2}+\frac{3}{2\pi}\hat\rho_{\Lambda}\ .
 \label{eq31d}
\end{equation} 
We can also substitute the density $\hat\rho_{\Lambda}$ by using Eqs.(\ref{eq30e}) and (\ref{eq31b}) and finally get
\begin{equation}
\frac{1}{y^2}\frac{d}{dy}\left(y^2\frac{dW}{dy}\right)=-W^{3/2}+\frac{2\rho_{\Lambda}}{\rho_p}\ ,
 \label{eq31e}
 \end{equation} 
which corresponds, if we take $\rho_p=\rho_{m0}$ and $W\equiv\theta$, exactly to the equilibrium equation for a polytropic configuration with index $n=3/2$ in presence of DE introduced by \cite{2012Merafina} in accordance to the dimensionless Emden variables and the initial conditions $W(0)=\theta(0)=1$ and $W'(0)=\theta'(0)=0$. Therefore, the polytropic configurations calculated by \cite{2012Merafina} in hydrostatic approach, can be used for the description of clusters of gravitating point masses with a distribution function (\ref{eq21b}) with energy cutoff (\ref{eq21c}), at small values of $\Lambda$ and large values of $T$.

\section{Numerical results}
The dimensionless equilibrium equation (\ref{eq29}) depends on two parameters: the gravitational potential at the center of the configuration $W_0$ and $\hat\rho_{\Lambda}$, which determines the intensity of DE through the value of the cosmological constant $\Lambda$. Different values of these parameters give a two-dimensional family of equilibrium solutions.
The set of solutions for $\hat\rho_{\Lambda} =0$ at different values of $W_0$ was obtained by \cite{1998Bisnovatyi}.

\begin{figure} \centering
\includegraphics[scale=0.5]{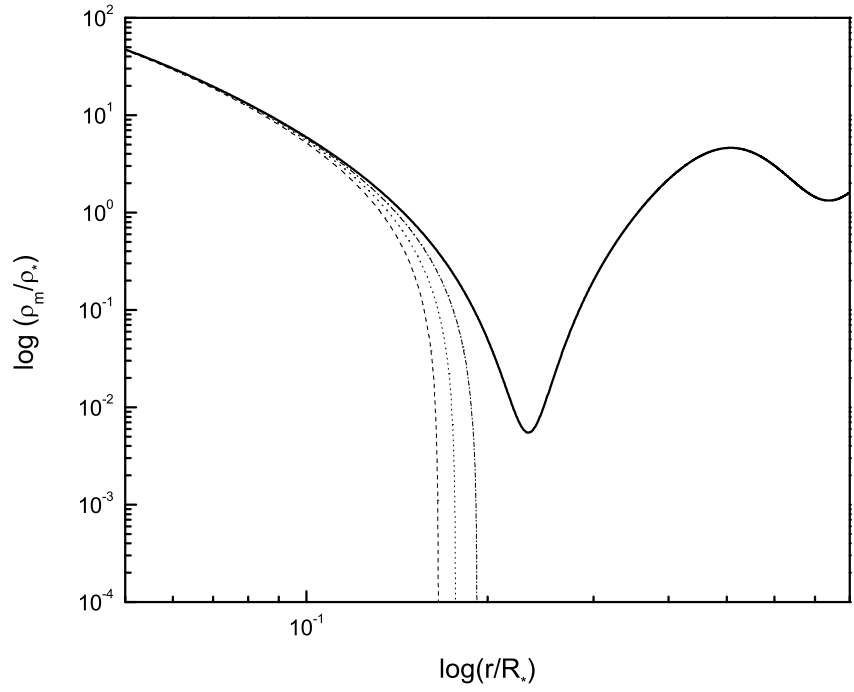}
\caption{Dimensionless matter density profiles for equilibrium configurations with $W_0=4$ and $\hat{\rho}_{\Lambda}=$ 0 (dashed line), 0.5 (dotted line), 0.9 (dash-dotted line), for which the matter density goes to zero, and $\hat{\rho}_{\Lambda}=$ 1.25 (solid line), chosen inside the region of non-equilibrium solutions.}
\label{fig1}
\end{figure}

\begin{figure} \centering
\includegraphics[scale=0.5]{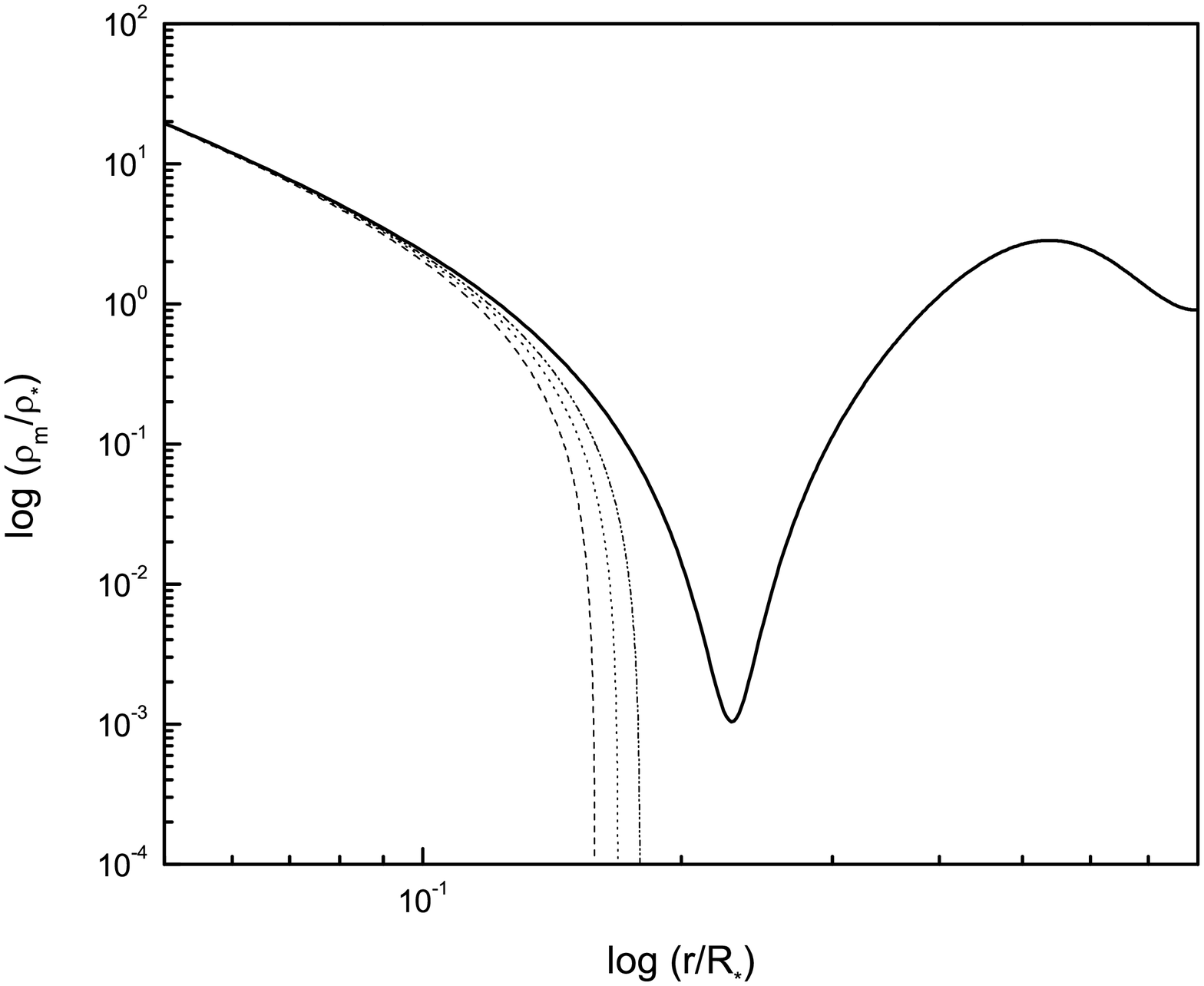}
\caption{Dimensionless matter density profiles for equilibrium configurations with $W_0=8$ and $\hat{\rho}_{\Lambda}=$ 0 (dashed line), 0.3 (dotted line), 0.5 (dash-dotted line), for which the matter density goes to zero, and $\hat{\rho}_{\Lambda}=$ 0.8 (solid line), chosen inside the region of non-equilibrium solutions.}
\label{fig2}
\end{figure}

\begin{figure} \centering
\includegraphics[scale=0.5]{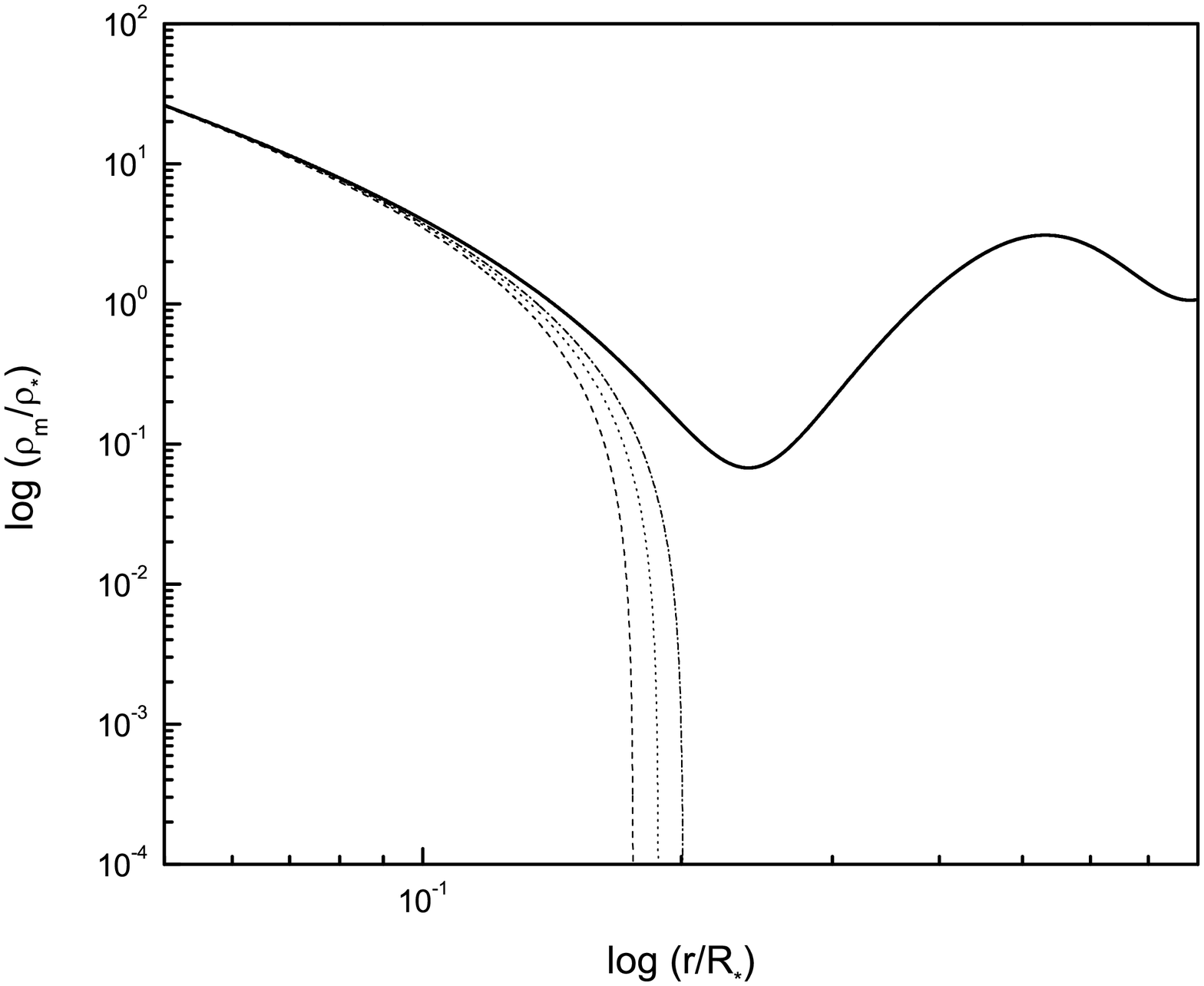}
\caption{Dimensionless matter density profiles for equilibrium configurations with $W_0=12$ and $\hat{\rho}_{\Lambda}=$ 0 (dashed line), 0.3 (dotted line), 0.5 (dash-dotted line), for which the matter density goes to zero, and $\hat{\rho}_{\Lambda}=$ 0.9 (solid line), chosen inside the region of non-equilibrium solutions.}
\label{fig3}
\end{figure}

We solved numerically the Poisson equation for gravitational equilibrium at different values of the two parameters ($W_0$, $\hat\rho_{\Lambda}$) mentioned above. First of all, we focused our attention on the matter density profiles $\rho_m(r)$ of the equilibrium configurations; in detail, we investigated how they change for increasing values of the dimensionless DE density $\hat{\rho}_{\Lambda}$ at fixed values of the dimensionless gravitational potential $W_0$. We chose three values of $W_0$ and, respectively, four values of $\hat{\rho}_{\Lambda}$, founding the existence of pairs of parameters ($W_0$, $\hat{\rho}_{\Lambda}$) which do not allow equilibrium solutions. This peculiarity is well represented in the matter density profiles shown in Figs.\,\ref{fig1}, \ref{fig2} and \ref{fig3}.
Hereafter in the figures, for the sake of compactness, we define the following quantities
\begin{equation}
\rho_* =\frac{\sigma^3}{G\zeta^2}\ , \quad M_*=\frac{\sigma^{3/2}\zeta}{G}\ , \quad R_*=\frac{\zeta}{\sigma^{1/2}}
 \label{eq37a}
\end{equation}
and, therefore, the dimensionless quantities introduced in Eqs.(\ref{eq30e}) and (\ref{eq30f}) can be rewritten as
\begin{equation}
\hat\rho_m =\frac{\rho_m}{\rho_*}\ ,\;\hat\rho_{\Lambda}=\frac{\rho_{\Lambda}}{\rho_*}\ ,\;\hat M^{(m)}=\frac{M^{(m)}}{M_*}\ ,\;\hat R=\frac{R}{R_*}\ ,\; \hat r=\frac{r}{R_*}\ .
 \label{eq37b}
\end{equation}
For each value of the central potential $W_0$, it exists one value of the 
parameter $\hat{\rho}_{\Lambda}$ after which the matter density profile does not converge to zero, but oscillates indefinitely. If we assert that the radius $R$ of an equilibrium configuration is defined as the value of the radial coordinate $r$ at which the matter density $\rho_m(r)$ becomes zero, it is clear that everytime this does not happen we are unable to estimate the radial extension of the system. All the configurations with a given value of $W_0$ and $\hat{\rho}_{\Lambda}$ which correspond to oscillating density profiles cannot be considered in gravitational equilibrium.

Moreover, the calculation of the total radius $R$ of an equilibrium configuration is strictly connected to the one related to the total mass $M^{(m)}$. Following Eq.(\ref{eq30c}) and expressing in terms of dimensionless quantities, the mass $\hat M_r^{(m)}$ within the radius $\hat r$ is given by
\begin{equation}
\hat M_r^{(m)}=\int_0^{\hat{r}} 4\pi \hat{\rho}_m \xi^2 d\xi \ .
\end{equation}
As a consequence, the non-equilibrium solutions for which the total radius $R$ cannot be defined do not even allow the evaluation of the total mass $M^{(m)}$ of the system.

\begin{figure} \centering
\includegraphics[scale=0.5]{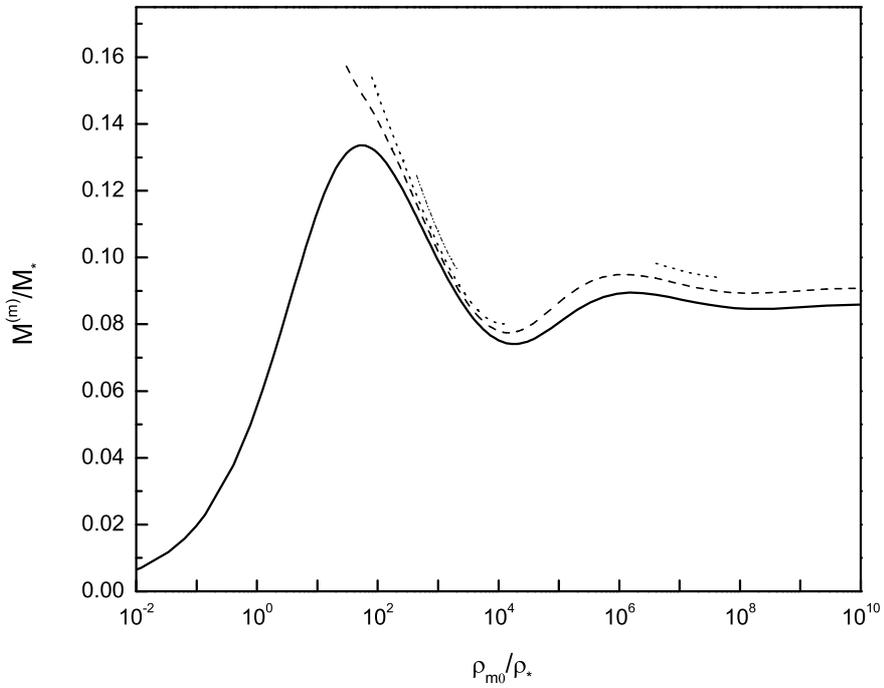}
\caption{Dimensionless mass as a function of the dimensionless central matter density for $\hat{\rho}_{\Lambda}=$ 0 (solid line), 0.5 (dashed line), 0.8 (dotted lines), 1.3 (dash-dotted line).}
\label{fig4}
\end{figure}

\begin{figure} \centering
\includegraphics[scale=0.5]{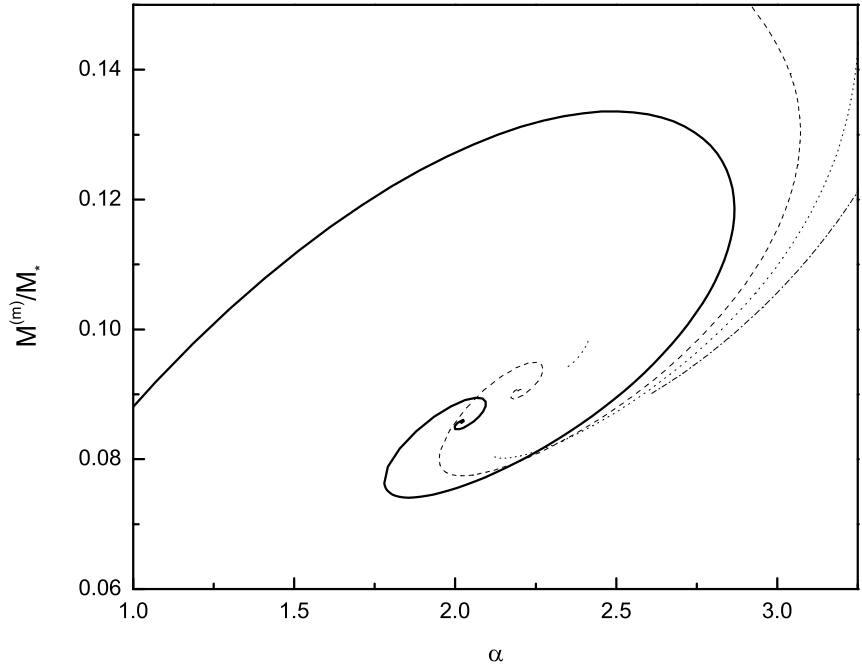}
\caption{Dimensionless mass as a function of the cutoff parameter $\alpha$ for $\hat{\rho}_{\Lambda}=$ 0 (solid line), 0.5 (dashed line), 0.8 (dotted lines), 1.3 (dash-dotted line).}
\label{fig5}
\end{figure}

\cite{1998Bisnovatyi} found the set of solutions at $\Lambda=0$ for $\hat{M}(\hat{\rho}_{m0})$ and $\hat{M}(\alpha)$ curves, in the Newtonian case.
These curves are shown in Figs.\,\ref{fig4} and \ref{fig5} (continuous line) together with the ones given for different values of $\hat{\rho}_{\Lambda}$ ($\Lambda\neq 0$). Analyzing the former figure (Fig.\,\ref{fig4}), when the parameter $\hat{\rho}_{\Lambda}$ is different from zero, and for increasing values of this parameter, the curves are not longer continuous and the absolute maximum of the mass disappears. Within the interval $0.6 \leq \hat{\rho}_{\Lambda}\leq 0.8$, the curves present several branches (in the figure, only the branches for $\hat{\rho}_{\Lambda}=0.8$ are shown in order to keep clear the understanding of the different behaviors). Out of the interval $0.6 \leq \hat\rho_{\Lambda}\leq 0.8$ the branches reduce to a unique curve and, in particular, for $\hat\rho_{\Lambda} > 0.8$, the curve becomes gradually shorter at increasing values of $\hat{\rho}_{\Lambda}$, until reaching the critical value $\hat{\rho}_{\Lambda}\simeq 1.38$ in which the curve reduces to a unique point (we discuss this critical value in the following). It is clear that the unusual behavior of the $\hat{M}(\hat{\rho}_{m0})$ is related to the density profiles of the non-equilibrium solutions. In order to analyze the latter figure (Fig.\,\ref{fig5}), by considering Eq.(\ref{eq30i}), we can conclude that also the parameter $\alpha$ is connected to the values of the total radius $\hat{R}$ and the mass $\hat M^{(m)}$. Consequently it is easy to show that for non-equilibrium solutions it is not possible to calculate the cutoff parameter $\alpha$. Therefore, also in this case we can expect the existence of different branches of solutions. Moreover, the behavior of the $\hat{M}(\alpha)$ curves at different values of $\hat\rho_{\Lambda}$ extends the range of solutions at values of $\alpha$ larger than the critical value ($\alpha=2.87$) valid for $\Lambda=0$ \citep{1998Bisnovatyi}. As previously underlined, it is possible to distinguish several branches of solutions with a limiting value of $\alpha$ changing in dependence of the value of $\hat\rho_{\Lambda}$. This limiting value, sistematically larger than 2.87, increases at increasing values of $\hat\rho_{\Lambda}$ until the absolute limiting value $\alpha_{lim}\simeq 3.42$.

The DE background in which all the bodies of the Universe are embedded, produces the antigravity that changes their gravitational equilibrium, acting on the contrary of the matter gravity. In order to establish when we are in presence of configurations where the presence of the DE can change the gravitational equilibrium, following \cite{2012Bisnovatyi}, we introduce the so called zero gravity radius $R_{\Lambda}$. This is a physical parameter which is defined as the distance from the center of the system where the matter gravity and DE antigravity balance each other exactly. 
Let us considering the total force acting on the unit mass
\begin{equation}
F=F_g+F_{\Lambda}=-\frac{GM^{(m)}_r}{r^2}+\frac{8\pi G\rho_{\Lambda}}{3}r\ ,
\label{eq38a}
\end{equation}
where, differently from Eq.(\ref{eq6}), this relation is also valid within the matter and not only in the vacuum. Then, the total force $F$ defined in Eq.(\ref{eq38a}) and, consequently, the acceleration, are both zero at the distance
\begin{equation}
R_{\Lambda}=\left[\frac{3M^{(m)}_{R_{\Lambda}}}{8\pi \rho_{\Lambda}}\right]^{1/3}\ ,
\label{eq38b}
\end{equation}
where the zero gravity radius depends on the total mass of the equilibrium configurations if $R\leq R_{\Lambda}$, while, if the condition $F=0$ is satisfied inside the configuration, we have not equilibrium and the mass to consider is $M_r^{(m)}$ with $r=R_{\Lambda}$. 

Therefore, every cluster has its zero gravity radius. This definition allows us to identify a gravitationally bound system only if it is enclosed within the sphere of radius $R_{\Lambda}$, namely only if its total radius is less than its zero gravity radius ($R<R_{\Lambda}$). Galaxies in the external regions where $r\ge R_{\Lambda}$ can flow out from the center of the cluster under the action of the DE antigravity force.

\begin{figure} \centering
\includegraphics[scale=0.5]{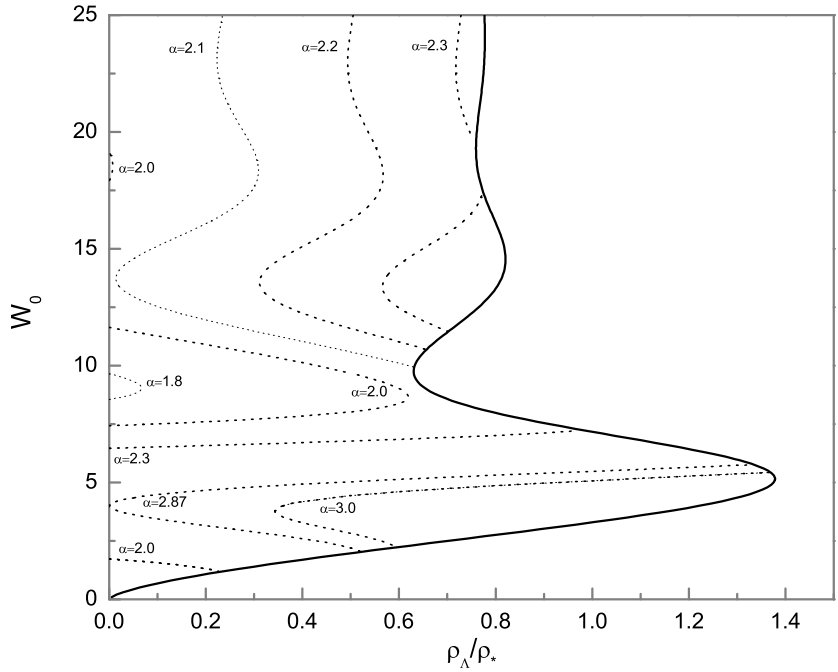}
\caption{Limiting curve of the equilibrium configurations with $R=R_{\Lambda}$ (solid line), expressed in terms of $W_0$ as a funtion of 
$\hat{\rho}_{\Lambda}$. Labelled curves at constant $\alpha$ (dashed lines) are also considered.}
\label{fig6}
\end{figure}

In Fig.\,\ref{fig6} we have represented the curve of the equilibrium configurations having $R=R_{\Lambda}$, through the behavior of $W_0$ as a function of $\hat{\rho}_{\Lambda}$. In addition we have also shown the curves representing the families of equilibrium solutions at fixed values of $\alpha$. By keeping constant the value of $W_0$ and increasing the value of 
$\hat{\rho}_{\Lambda}$ there exists one limiting value, lying on the curve, after which it is no longer possible to obtain equilibrium solutions. Along this limiting curve, which separates two regions (solid line), the equilibrium configurations have the total radius exactly equal to the zero gravity radius and the matter density profiles vanishing with a minimum in correspondence of the total radius $R=R_{\Lambda}$. Therefore, it is possible to define the region on the right side of the figure in which no gravitational equilibrium can establish and no curves at constant 
$\alpha$ can lie, corresponding to configurations with matter density profiles not converging to zero. On the contrary, in the region corresponding to the left side of the figure, we can assert that the force due to the presence of the DE, $F_{\Lambda}$, is less strong than the one due to the gravity, $F_g$, and the gravitational equilibrium can be achieved. In other words, speaking in terms of radial extension, the condition $R\le R_{\Lambda}$ is satisfied for each configuration belonging to this region and the matter density profiles are regular and converging to zero in correspondence of the total radius $R$. 

Finally, from Eq.(\ref{eq38b}), we can see that the zero gravity radius $R_{\Lambda}$ is inversely proportional to the DE density $\rho_{\Lambda}$. Therefore, by decreasing the value of the DE density, the zero gravity radius increases until the condition $R_{\Lambda}\to\infty$ for 
$\rho_{\Lambda}=0$. Actually, by considering the plane ($W_0$-$\hat{\rho}_{\Lambda}$) of Fig.\,\ref{fig6}, the gravitational equilibrium is achieved more easily, and for more values of $W_0$, when the $\hat{\rho}_{\Lambda}$ parameter is small. In particular, for $\hat{\rho}_{\Lambda}=0$, the equilibrium solutions are possible for each value of $W_0$, recovering the well know results of \cite{1998Bisnovatyi}.

\section{Conclusions}

We have calculated the equilibrium configurations of Newtonian clusters with a truncated Maxwellian distribution function, in presence of DE. All these clusters have a structural equilibrium, being satisfied the condition $R<R_{\Lambda}$, and result dynamically stable. By considering the relaxation time, we obtain a value larger than the age of the Universe and, therefore, we can conclude that thermodynamical instabilities are not relevant in current evolution of galaxy clusters. On the other hand, the evaluation of parameters characterizing this kind of clusters suggests that these systems are collisionless. In any case, the critical point of the onset of thermodynamic instability lies far from the first maximum mass, at larger values of $W_0$ in the curve with $\hat{\rho}_{\Lambda}=0$ of Fig.\,\ref{fig4} \citep{2006bkm}, as well in curves with $\hat{\rho}_{\Lambda}\neq 0$, the most part of the equilibrium configurations results thermodynamically stable.

Presently the density distribution inside galaxy clusters is described by several phenomenological functions, some if which follow from numerical simulations \citep[see][]{2013chernin}. Qualitatively the truncated Maxwellian distribution, considered here is similar to the non-singular density distribution suggested by \cite{2013chernin}. It may be used for the more detailed study of the density and velocity  distributions on the periphery of rich clusters, where the influence of DE is important, and their comparison with observations. The galaxies in the outer parts of the clusters are not numerous, and they have smaller masses and luminosities in presence of even weaker relaxation. Therefore, only largest telescopes should be used  for a search of galaxies at the cluster peripheries.  The most sensitive X-ray telescopes are needed for detection of the hot gas in the outer parts of the clusters, and its possible outflow in presence of DE, considered by \cite{2013bkm}. 

\section*{Acknowledgements}

The work of G.S.\,Bisnovatyi-Kogan was partly supported by RFBR grant 11-02-00602, the RAN Program Formation and evolution of stars and galaxies, and Russian Federation President Grant for Support of Leading Scientific Schools NSh-5440.2012.2\,.


\bibliographystyle{plainnat}

\bibliography{biblio}




\onecolumn

\end{document}